\documentclass[aps,twocolumn,superscriptaddress,amsmath,amsfonts,showkeys,
               floatfix]{revtex4}

\usepackage{epsfig,psfrag}
\usepackage{CJK}

\usepackage{graphicx}
\usepackage{dcolumn}
\usepackage{bm}
\usepackage{amssymb}
\usepackage{natbib}

\begin{document}
\title{Mobile defects as mediated states for charge-carrier trapping in metal halide perovskites quantum dots}
\author{Xiao-Yi Liu}
\affiliation{Tianjin Key Laboratory of Low Dimensional Materials Physics and Preparing Technology, Department of Physics, School of Science, Tianjin University, Tianjin 300354 China}
\author{Wei-Ping Li}
\affiliation{Department of Basic Courses, Tianjin Sino-German University of Applied Sciences, Tianjin 300350, China}
\email{lwpliweiping@126.com}
\author{Yu Cui}
\affiliation{Tianjin Key Laboratory of Low Dimensional Materials Physics and Preparing Technology, Department of Physics, School of Science, Tianjin University, Tianjin 300354 China}
\author{Shao-Juan Li}
\affiliation{State Key Laboratory of Luminescence and Applications, Changchun Institute of Optics, Fine Mechanics and Physics Chinese Academy of Sciences, Changchun 130033, China}
\author{Ran-Bo Yang}
\affiliation{Tianjin Key Laboratory of Low Dimensional Materials Physics and Preparing Technology, Department of Physics, School of Science, Tianjin University, Tianjin 300354 China}
\author{Zhi-Qing Li}
\affiliation{Tianjin Key Laboratory of Low Dimensional Materials Physics and Preparing Technology, Department of Physics, School of Science, Tianjin University, Tianjin 300354 China}
\author{Zi-Wu Wang}
\email{wangziwu@tju.edu.cn}
\affiliation{Tianjin Key Laboratory of Low Dimensional Materials Physics and Preparing Technology, Department of Physics, School of Science, Tianjin University, Tianjin 300354 China}

\begin{abstract}
The migration motion of defects in metal halide perovskites quantum dots (MHPQDs) results in charge-carrier trapping become more complicated. We study two-step trapping mediated by mobile defects between the ground state of MHPQDs and a fixed-depth defect using a full-configuration defect method, where all possible trapping processes mediated by these mobile defects could be reproduced and the fastest channels among them are picked out. We find that these two-step trapping processes could keep more one order of magnitude faster than these direct ones as mobile defect with the appropriate localization strength, which implies that these indirect trapping should play the crucial rule to determine the non-radiative recombination losses. These results provide the significant explanation for studying non-radiation processes of carriers in the presence of the migration defects in recent experiments. Moreover, this model will be available to analyze some key performance related defects in electronic devices.
\end{abstract}
\maketitle

The exceptional properties of metal halide perovskites quantum dots (MHPQDs) have aroused extensive attention in the past decade owing to their potential applications in the next-generation of photovoltaic and photoelectric devices\cite{R1,R2,R3}. Various defects in these nanostructures that arising from the solution-fabrication processes and the original soft lattice, paly a vital role to determine the devices performance\cite{R4,R5,R6,R7,R8,R9}, even though the property of "highly defect tolerance" has been widely accepted for most metal halide perovskites materials\cite{R10,R11,R12,R13,R14}. One of the pernicious impacts of these defects is that acting as trapping centers result in the inevitable non-radiative recombination losses and thus cause the changing of some key parameters of device performance, such as open-circuit voltage\cite{R15,R16}, short-circuit current density\cite{R17,R18} and power-conversion efficiency for perovskite solar cells\cite{R16,R17}. In particular, carriers trapping become more complicated in the presence of a large amount of migration defects in these perovskites nanostructures\cite{R19,R20,R21,R22,R23,A1}.
\begin{figure}
\includegraphics[width=3.45in,keepaspectratio]{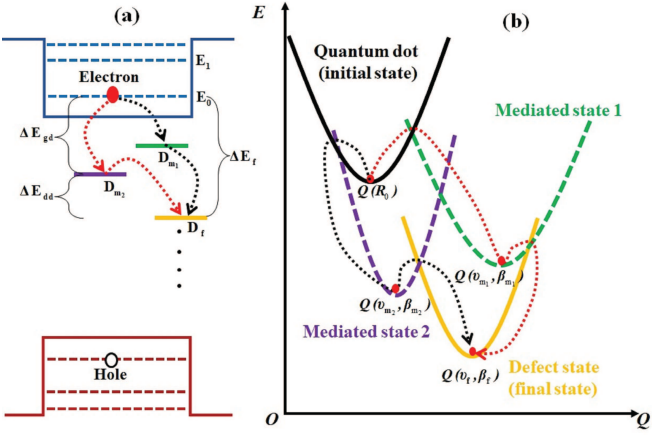}
\caption{\label{compare} (a) the schematic diagram of carrier trapping between the ground state of MHPQDs ($E_0$) and a fixed-depth defect ($D_f$) mediated by mobile defects ($D_{m1}$, $D_{m2}$$\cdots$) in the bandgap representation. $\Delta E_{gd}$ ($\Delta E_{f}$) and $\Delta E_{dd}$ denote the energy separation between the ground state of MHPQDs and the mobile defect (the fixed-depth defect) as well as between the mobile defect and the fixed depth defect, respectively. (b) the schematic diagram of the trapping processes in the lattice coordinate configuration, where $Q(R_0)$ is the equilibrium coordinate of the ground state of MHPQDs depending on the radius of quantum dot $R_0$; $Q(\nu_i,\beta_i)$ denotes the equilibrium coordinate of defect with two parameters $\nu$ and $\beta$ describing the depth and the different lattice relaxation strength, respectively.}
\end{figure}

In fact, there have been an abundance of studies for charge carriers in the ground states of MHPQDs (or the edges of band-gap) are trapped by defects\cite{R4,R5,R6,R7,R8,R9,R16,R17,R18,R19} as well as the trapping processes between two defects with different depthes\cite{R7,R18,A2}. From these studies, one can conclude that charge-carriers trapping are hindered substantially with increasing defect depth, because these trapping processes mediated by deeper defects need more phonon numbers to match the energy difference between defects and ground state of MHPQDs. However, the migration of some defects that acting as the efficient mobile trapping centers accelerate non-radiative recombination by an order of magnitude, because the migration induces lattice distortion, providing the enhanced lattice relaxation energy (more phonon numbers) are needed in trapping processes\cite{R21,R22,R23,A1}. These mobile defects with unfixed depth and variable electrical polarizability of the lattice around them, on the one hand, enhance charge carriers trapping significantly. On the other hand, amounts of possible trapping channels induced by mobile defects bring more challenge for analyzing carriers trapping processes, even though the great computational labor could be undertaken for more accurately simulations in first-principles calculations\cite{R8,R9,R12,R20}. In order to overcome this problem, we developed a method of full-configuration defect\cite{A3} to describe the key feature of mobile defect that the variable depth with different lattice relaxation strength based on the classical quantum defect model\cite{R24,R25,R26}, where  all possible channels of charge-carrier trapping by mobile defects could be given. Although so many researches have been focused on these mobile defects, studies for mobile defects serving as the bridge states to assist charge-carrier transfer between the ground state of MHPQDs and the deep defects are still lacking until now.

In the present paper, we study two-step trapping processes by using a full-configuration defect method, in which an electron is trapped by a mobile defect from the ground state of MHPQDs firstly, then this electron transfers from the mobile defect to the fixed-depth defect via non-radiative multiphonon processes. With the aid of this method, the fastest channel could be picked out from all possible two-step processes of charge-carrier trapping mediated by these mobile defects. We find that although these trapping processes are indirect, the trapping time is more faster than these direct processes as mobile defect with an appropriate localization parameter, which indicates that these two-step trapping processes should coexist with these direct ones in MHPQDs, playing the crucial rule to determine the non-radiative recombination losses. These numerical results not only enrich the knowledge for analyzing non-radiative recombination in MHPQDs, but provide the potential explanation for some unique properties of defects observed by recent experiments. More importantly, this two-step trapping could be expanded to analyze some key performances of electronic devices, such as metastable states of defects resulting in the bias temperature instability for devices\cite{R27,R28,A4} and Franck-Condon blockade in single molecular junction\cite{R29}.

As schemed in Fig. 1, the whole trapping process is divided into two steps: (i) an electron in the ground state of quantum dot is trapped by defects; (ii) electron transfer from the mobile defects to the fixed-depth defects. Within the frame of Huang-Rhys model\cite{R30,R31}, the non-radiative recombination via multiphonon processes for two steps could be, respectively, expressed as
\begin{eqnarray}
\tau_{gd}^{-1}&=&\frac{2\pi|H_{gd}|^2}{\hbar(k\hbar\omega_{LO})}\left(\frac{\overline{n}_{LO}+1}{\overline{n}_{LO}}\right)^{\frac{k}{2}}\exp\left[-S_{gd}(2\overline{n}_{LO}+1)\right]\nonumber\\
&&I_k\left[ 2S_{gd}\sqrt{\overline{n}_{LO}(\overline{n}_{LO}+1)}\right],
\end{eqnarray}
\begin{eqnarray}
\tau_{dd}^{-1}&=&\frac{2\pi|H_{dd}|^2}{\hbar(p\hbar\omega_{LO})}\left(\frac{\overline{n}_{LO}+1}{\overline{n}_{LO}}\right)^{\frac{p}{2}}\exp\left[-S_{dd}(2\overline{n}_{LO}+1)\right]\nonumber\\
&&I_p\left[ 2S_{dd}\sqrt{\overline{n}_{LO}(\overline{n}_{LO}+1)}\right],
\end{eqnarray}
where $H_{gd}$ ($H_{dd}$) denotes the transition matrix between the ground state of MHPQDs and a mobile defect (between a mobile defect and a fixed-defect) induced by electron-longitudinal optical (LO) phonon interaction. The detailed expressions for $H_{gd}$ and $H_{dd}$ are given in the supplementary materials. $S_{gd}$ and $S_{dd}$ are the well-known Huang-Rhys factors (whose formulations are shown in the supplementary materials), describing the difference of lattice relaxation strength between the ground state of MHPQDs and a mobile defect as well as between two defects, respectively. See the supplementary materials for their relations with the depth of mobile defect in Fig. S1. Except for the depth of defect, the mobile feature of defects can be reflected in both transition matrixes and Huang-Rhys factors by the localization parameter $\beta_m$ that denotes defect-LO phonon coupling strength, see Eqs. (S20), (S21), (S28) and (S29) in the supplementary materials. The energy difference $\Delta E_{gd}$ between the ground state of MHPQDs and a mobile defect is expressed in units of LO phonon energy $\hbar\omega_{LO}$, satisfying $\Delta E_{gd}$=$k$$\hbar\omega_{LO}$ ($k$=1,2,3$\cdots$). Similarly, the energy difference $\Delta E_{dd}$ between two defects follows the relation of $\Delta E_{dd}$=$p$$\hbar\omega_{LO}$ ($p$=1,2,3$\cdots$). $\bar{n}_{LO}$ denotes the thermal average LO phonon number, and $I_k$ ($I_p$) is the modified Bessel functions of the first kind. Here, LO phonon modes are mainly considered because they give the dominate contribution to carrier-phonon scattering proved by experiments extensively\cite{R32,R33,R34}. In the numerical calculation, neutral defects are selected and some related parameters are listed in Table SI of supplementary materials.
\begin{figure}
\includegraphics[width=3.4in,keepaspectratio]{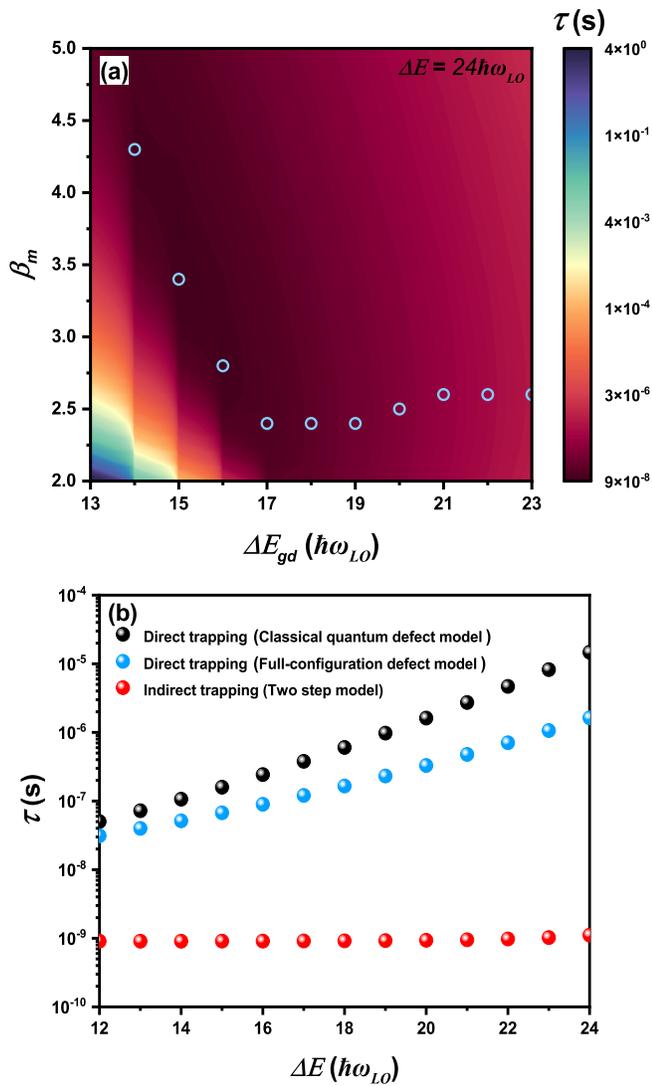}
\caption{\label{compare} (a) The two-step process of the trapping time $\tau$ as functions of the variable depth of mobile defects $\Delta E_{gd}$ and the polarization parameter $\beta_m$ for the fixed defect depth $\Delta E_{f}$=24$\hbar\omega_{LO}$ and temperature $T$=300 K, in which these minimal values of $\tau$ (corresponding to the fastest relaxation channel) are signed by circles; (b) The trapping time as a function of the fixed-defect depth for an electron trapping from the ground state of quantum dot to the fixed-depth defect at  temperature $T$=300 K, where the comparisons of the trapping time for three model are shown: the classical quantum defect model (black circles), the full-configuration defect model (blue circles) and the mobile defect-mediated model (res circles). }
\end{figure}

According to Eqs. (1) and (2), the total trapping time of the indirect process is the summation of two steps $\tau$=$\tau_{gd}$+$\tau_{dd}$. In Fig. 2 (a), the total lifetime (TLT) for overall possible channels for an electron trapping from the ground states of quantum dot to a fixed-depth defect mediated by mobile defects with variable depth ($\Delta E_{gd}$) and localization parameter ($\beta_m$) are presented. One can see that TLT can vary in a large scale depending on the localization parameter sensitively when the level of mobile defect is not very close to the fixed-depth defect, which displays that trapping abilities are different for the same-depth defects due to their different lattice relaxation strength. More importantly, these minimal values of TLT that correspond to the fastest channel of electron trapping could be singled out as given by these circle symbols, which shows that this two-step trapping process is very fast, maintaining in the scale of tens $\sim$ hundreds of nanoseconds in the presence of mobile defects with the appropriate localization parameters, which is very consistent with the experimental measurements\cite{R4,R16,R18,R21}. In Fig. 2 (b), the fastest process for an electron is trapped from the ground state of quantum dot to a defect are compared for three models: the classical quantum defect model, the full-configuration defect model and the mobile defect-mediated model (or the two-step model). From the comparisons, one can see that the two-step trapping process keeps more one order of magnitude faster than the former two direct models, which strongly suggest that this type of the indirect process should play a very important rule to determine non-radiative recombination losses, thus the power-conversion efficiency in photovoltaic and photoelectric devices\cite{R1,R2,R3,R4,R16,R18}.

From Eqs. (1) and (2),  one can conclude that each transition in this two-step process depends on the temperature described by the average LO phonon number $\bar{n}_{LO}$,  both of which, in fact, are thermally excited processes in the classical Huang-Rhys model\cite{A2,A3,R25,R26,R30,R31}. The temperature dependence of the total trapping time with different localization parameters are plotted in Fig. 3. The trapping time decreases one order of magnitude with temperature, because more LO phonons will be excited with increasing temperature, thus benefiting to these thermally excited trapping processes. The most fast trapping channels for these two-step processes with the optimal localization parameters are also signed by these circle symbols in Fig. 3, which show that the optimal mediated-defects will be shifted to defects with smaller $\beta_m$ when the temperature increases. In addition, the electron trapping in the first-step process depends on the radius of quantum dot due to the quantum confinement effect, which has been discussed in our previous study\cite{A3}.

Apart from providing the explanation for non-radiative recombination losses, we notice that this two-step process may be available to analyze many performances of electronic devices related to charge-carrier trapping or transfer by defects. For instance, in the past decades, in order to give the reasonable explanation for the unusual behaviors of some defects in field-effect transistors by the time-dependent defect spectroscopy, Grasser et al. proposed a multi-state model for charge-carrier trapping by defects, in which the metastable states of defects are introduced and occupied temporarily, serving as the mediated states for carriers transfer\cite{R27,R28,A4}. These metastable states of defects could be completely reproduced by the mobile feature of defects as suggested in the present model. In a recently experiment, the sequential tunneling and cotunneling of electron transfer can be tuned between the right and the left graphene electrodes in a single-molecule junction due to the strong electron-phonon coupling\cite{R29}, where the tunable mediated-states of electron-transfer induced by the torsional molecular motions are very similar to the mobile defect states. Therefore, these theoretical results also provide the enlightenment for studying quantum transitions of charge-carriers between nanoscopic system and its environment.
\begin{figure}
\centering
\includegraphics[width=3.3in,keepaspectratio]{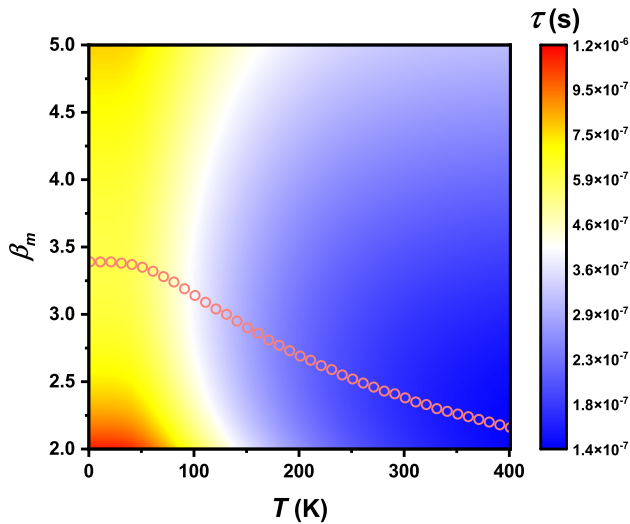}
\caption{The total trapping time of two-step trapping process as functions of the temperature and the localization parameter ($\beta_m$) for the neutral defects at $\Delta E_{f}$=24$\hbar\omega_{LO}$ and $R_0$=40 nm. The circle symbols denote the shortest trapping time (or the fastest trapping channel) for mobile defects with the appropriate localization parameters.}
\end{figure}

In summary, we investigate the charge-carrier trapping between the ground states of quantum dot and a fixed depth defects mediated by mobile defects based on the full-configuration defects method. We find that (1) the total trapping time of this two-step process vary from tens of to hundreds of nanoseconds, depending on the depth and the localization strength of mobile defects; (2) this two-step trapping process could keep more one order of magnitude faster than these direct processes as the mobile defects with the appropriate localization strength, which implies that this two-step process will play an important role to determine the dynamic properties of charge-carrier transfer in nanostructures. Moreover, this type of process could be expanded to analyze some key performances of devices related to carriers trapping by defects directly and indirectly. We hope these results can stimulate more and more future experiments in this aspect.

See the supporting information for the eigenfunctions and eigenenergies of an electron in a quantum dot, the full-configuration quantum defect model, the calculation of Huang-Rhys factor and the detailed derivations for the trapping probability of charge-carrier by defects.

This work was supported by National Natural Science Foundation of China (Nos 12174283,  62022081 and 61974099).

The data that support the findings of this study are available from the corresponding author upon reasonable request.

\end{document}